\documentclass{iopart}

\usepackage{graphicx}
\usepackage{dcolumn}
\usepackage{bm}
\usepackage{epsfig}
\usepackage{iopams}

\begin{document}

\title[Spin resolved energy parametrization of a quasi-one-dimensional electron gas]{Spin resolved energy parametrization of a quasi-one-dimensional electron gas}

\author{Luke Shulenburger,$^{1}$ Michele Casula,$^{2}$ Gaetano
Senatore,$^{3,4}$ and Richard M. Martin$^{5}$}

\address{ $^1$ Carnegie Institution of Washington, 5251 Broad Branch Road, NW, Washington, D. C.
20015, USA\\
$^2$ Centre de Physique Th\'eorique, Ecole Polytechnique, CNRS, 91128 Palaiseau, France \\
$^3$ INFM-CNR Democritos National Simulation Center, Trieste, Italy \\
$^4$ Dipartimento di Fisica Teorica dell' Universit\`a di Trieste, Strada
Costiera 11, 34014 Trieste, Italy \\
$^5$ Stanford University Physics Department, 382 Via Pueblo Mall 
Stanford, CA 94305, USA }

\ead{michele.casula@gmail.com}

\date{\today}

\begin{abstract}
By carrying out extensive lattice regularized diffusion Monte Carlo calculations,
we study the spin and density dependence of the ground state energy for a quasi-one-dimensional electron gas,
with harmonic transverse confinement and long-range $1/r$ interactions. We present a parametrization
of the exchange-correlation energy suitable for spin density functional calculations, which fulfills
exact low and high density limits. 
\end{abstract}



\section{Introduction}
In this paper we present a parametrization for the exchange-correlation energy of a
quasi-one-dimensional electron gas (1DEG) at arbitrary polarization.
The electrons interact via a $\frac{1}{r}$ potential and are confined
to a line by a transverse harmonic potential $v(r_\perp) = \frac{r_\perp^2}{4b^4}$,
where $b$ controls the thickness of the wire.
Here and henceforth we use the effective Bohr radius
$a^\star_0 = \frac{\hbar^2 \epsilon}{m^\star e^2}$ as unit of length and the
effective Rydberg $Ryd^\star = \frac{e^2}{2\epsilon a_0^\star}$ as unit of energy,
where $\epsilon$ is the dielectric constant of the embedding medium and
$m^\star$ is the effective electron mass.
We can separate the transverse and longitudinal
parts of the Hamiltonian by assuming that the electrons are in the ground
state of the 2D harmonic oscillator in the transverse direction. This is a
good approximation provided that $r_s \gg \frac{\pi b}{4}$,
where $r_s$ is the Wigner-Seitz radius.
The above condition is met for low enough densities and thin enough wires.
Thus, it is possible to integrate out
the perpendicular degrees of freedom and work with a strictly
one-dimensional Hamiltonian with electrons interacting via an effective
potential given by
\begin{equation}
V_b(x)= \frac{\sqrt{\pi}}{b} {\rm exp} \left( \frac{x^2}{4 b^2} \right) {\rm
  erfc} \left(\frac{ |x|}{2 b} \right).\label{potential}
\end{equation}
This potential has been widely used in previous works to model the 1DEG, and we refer the
reader to Ref.~\cite{casula-1d} for a detailed description of the
Hamiltonian we study and the variational wave function we use.
Here, we employed the lattice regularized diffusion Monte Carlo (LRDMC)\cite{casula_lrdmc} algorithm to
compute the ground state energy of the system at different densities $r_s$ and polarizations,
($\zeta\equiv(N^\uparrow-N^\downarrow)/N)$. 
In one dimension, this method provides the exact energy within the statistical accuracy,
since the nodes of the ground state wave function are known exactly.

Despite the huge amount of work done for 1D systems with a $1/r$ interaction,
\cite{casula-1d,shulenburger,gold-excit,gold-gs-osci,gold-calmels-spin-susc,gold-hf,gold-rpa-msa}
a spin density exchange-correlation functional is still lacking, and
the Bethe ansatz solution is not available in this case. Fogler\cite{fogler-coulomb-tonks-prl}
derived an approximate mapping of the problem with a realistic Coulomb interaction
onto exactly solvable models of mathematical physics, but the relation is valid only
for ultra-thin wires and requires a careful matching between different regimes.
The quantum Monte Carlo framework can provide a parametrization valid in all regimes,
but so far a functional has been derived only for an unpolarized wire.\cite{casula-1d}
The present work fills this gap, and we provide a spin dependent density functional
for the exchange and correlation energy suitable for DFT calculations of these systems.
Indeed, the DFT framework has been applied quite successfully in 1D,
\cite{magyar,polini1,polini2,polini3} mainly on short-range 1D problems where the
homogeneous reference was known via Bethe ansatz.

The paper is organized as follows.
In Sec.~\ref{parametrization} we show the results for the
ground state energy and give a parametrization for
the exchange-correlation part, while in Sec.~\ref{conclusions} we present conclusions.
In the Appendix we derive the
polarization dependent random phase approximation (RPA) expression for the correlation energy,
which is used to set the high density limit of our parametrization.

\section{Exchange-Correlation Energy and Construction of an LSDA Functional}
\label{parametrization}
We study the ground state energy of the 1DEG
as a function of density and spin polarization, and find
a parametrization for the exchange-correlation energy
based on theoretically known properties of the electron gas in various limits.
The best parameters for the exchange-correlation functional will be
determined via a $\chi^2$ minimization of our LRDMC values
for the total energy.

Following the usual notation, we separate the total energy $\epsilon$ into three parts:
\begin{equation}
\epsilon(r_s,\zeta) = \epsilon_{t}(r_s,\zeta) +
\epsilon_{x}(r_s,\zeta) + \epsilon_{c}(r_s,\zeta)
\end{equation}
where $\epsilon_t$ is the kinetic energy of the noninteracting system,
$\epsilon_x$ is the
exchange energy calculated for the noninteracting wavefunction and
$\epsilon_c$ is the correlation energy which includes corrections to both
the potential energy and also the kinetic energy due to the interactions.
The first two terms are known analytically, while the third one is
fully determined by our numerical results for the total energy $\epsilon$.
The kinetic energy reads
\begin{equation}
\label{one_body_kinetic_energy}
\epsilon_t(r_s,\zeta) = \frac{\pi^2(1 + 3\zeta^2)}{48r_s^2},
\end{equation}
while the exchange energy is
\begin{eqnarray}
\label{hartree-fock-exchange}
\epsilon_x(r_s,\zeta) =& \frac{1 + \zeta}{2 b} ~~
F\left(\frac{4 r_s}{(1 +\zeta)\pi b}\right) + 
\frac{1-\zeta}{2 b} ~~ F\left(\frac{4 r_s}{(1 -\zeta) \pi b}\right) \\
\nonumber
F(x) =& -\int_0^{2/x} dy \tilde{v}(y) \frac{1 - xy/2}{2\pi},
\end{eqnarray}
with $\tilde{v}(x) = 2 E_1(x^2) \exp(x^2)$ the Fourier transform of the potential
in (\ref{potential}), where $E_1$ is the exponential integral function.

To derive an accurate parametrization for the
exchange-correlation energy, it is useful to
study both the high and low density limits in order to
include them in the actual functional. The high density
limit is estimated with the random phase approximation (RPA),
while the low density physics is obtained through a mapping onto
an effective one dimensional Heisenberg model. 

The RPA is very successful in describing the
energy of the homogeneous electron gas at high
density\cite{gold-gs-osci,casula-1d}.
Here we present the main result valid for $r_s \ll 1$
with the effective interaction in (\ref{potential}),
while a detailed derivation
is given in the \ref{RPA_app}. 
It is worth stressing that in the high density limit (small $r_s$) 
the 1D model with effective pair interactions given by (1) 
does not accurately describe electrons confined in a transverse harmonic potential,
since the condition $r_s \gg \frac{\pi b}{4}$ is in general manifestly violated,
and the single subband approximation breaks down.
The correlation energy evaluated within the RPA is
\begin{equation}
\epsilon^{RPA}_{c}(r_s,\zeta) = \left \{
\begin{array}{ll}
- C\left(1+\frac{1}{1-\zeta^2}\right) r_s^2  & \textrm{if $ r_s\ll (1-\zeta) \pi b/2$} \\
 - \frac{C}{4} r_s^2 & \textrm{if $\zeta=1$},
\end{array}
\right.
\label{RPA_full1}
\end{equation}
where $C = \int_0^\infty\, z \tilde{v}^2(z) dz/(2 \pi^4 b^2) \approx 4.9348/(2 \pi^4 b^2)$.
Though from (\ref{RPA_full1}) the correlation energy may at first appear discontinuous 
at $\zeta=1$, $\epsilon^{RPA}_{c}(r_s,\zeta)$ is in fact a continuous function 
of its variables as the two limiting behaviours in (\ref{RPA_full1}) 
clearly belong to different regions in the $\zeta,r_s$ plane.  

The low density dependence of the correlation energy is difficult to determine
since the effective coupling is very strong. This causes the electrons
to repel each other and form a quasi-Wigner crystal.\cite{shulenburger}
As the exchange between the particles drops off very rapidly with the reduction in the density,
different spin configurations become almost degenerate. However,
the Lieb-Mattis theorem\cite{mattis-lieb-ll}
proves that in one dimension the ground state  energy of a system of
fermions corresponds to zero total spin.
This theorem precludes the existence of a Bloch instability such as that
predicted by an STLS-like theory.\cite{gold-gs-osci,gold-calmels-spin-susc}
The low density spin dependence of the correlation energy
can be determined
approximately by noting that the spin sector of the 1DEG can be mapped to that of a Heisenberg spin
chain\cite{giamarchi-book} with coupling $J$.  In fact at these densities the
electron gas is a quasi-Wigner crystal with local antiferromagnetic correlations.
\cite{hausler-qmc-quantum-wires,casula-1d,shulenburger}
The Heisenberg coupling can be determined by
an evaluation of the tunneling (exchange) rate between electrons via
the WKB approximation, which gives
an exponential suppression of $J$ at low density
as stated by the relation:\cite{matveev-conductance-of-quantum-wire}
\begin{equation}
J(r_s) = \frac{J^{\star}}{(2r_s)^{1.25}} e^{-\nu \sqrt{2 r_s}}
\label{jwkb}
\end{equation}
where $J^{\star}$ and $\nu$ are interaction dependent constants.
The energy dependence as function of $J$ of the antiferromagnetic Heisenberg spin chain
 is known exactly from the Bethe ansatz.\cite{griffiths-magnetization-heisenberg}
The difference in energies between the polarized and unpolarized spin chains turns out to be
$J \ln2$.\cite{hulthen} These relations define the spin dependence of
the total energy of the electron gas at low density. Note that
in order to provide the
exponentially small spin dependence given by (\ref{jwkb}),
the correlation energy must cancel the power law and
logarithmic terms of both the exchange and kinetic terms.

Our spin dependent exchange-correlation functional is built upon the parametrization
of the exchange and correlation energy
for the unpolarized ($\zeta=0$) and polarized ($\zeta=1$) wires,
which reads:
\begin{equation}
\epsilon_{xc}(r_s,\zeta) = \frac{a_\zeta + b_\zeta r_s + c_\zeta r_s^2}{1 + d_\zeta  r_s +
e_\zeta  r_s^2 + f_\zeta r_s^3} + \frac{g_\zeta  r_s  \ln \left[r_s + \alpha_\zeta r_s^{\beta_\zeta}\right]}{1 +
h_\zeta r_s^2},
\label{exc_zeta}
\end{equation}
where the parameters are constrained to fulfill the high density limits of both exchange and correlation terms.
Those limits imply the following conditions on the parameters:
\begin{eqnarray}
a_0 & =  -\frac{\sqrt{\pi}}{2b}, \\
a_1 & =  -\frac{\sqrt{\pi}}{2b}, \\
b_0 & =  \frac{2+\gamma+2 \ln(\pi b/2)}{\pi^2 b^2} + a_0 d_0, \\
b_1 & =  \frac{2+\gamma+2 \ln(\pi b)}{2 \pi^2 b^2} + a_1 d_1, \\
c_0 & =  -2C + \frac{2+\gamma+2 \ln(\pi b /2)}{\pi^2 b^2} d_0  + a_0 e_0, \\
c_1 & =  -C/4 + \frac{2+\gamma+2 \ln(\pi b)}{2 \pi^2 b^2} d_1 + a_1 e_1, \\
g_0 & =  -\frac{2}{\pi^2 b^2}  \textrm{~~~with $\beta_0>1$}, \\
g_1 & =  -\frac{1}{\pi^2 b^2}  \textrm{~~~with $\beta_1>1$},
\end{eqnarray}
where $\gamma=0.5772156649$ is the Euler's constant.
On the other hand, the large $r_s$ expansion of the expression in (\ref{exc_zeta}) goes as $\ln{r_s}/r_s$.
Indeed, in previous work\cite{gold-gs-osci,casula-1d} it was found that both
the correlation and the exchange energies go as $\ln{r_s}/r_s$ at large $r_s$,
with their ratio approaching a constant in that limit,
a condition which is fulfilled by our parametrization.

It is useful also to define a constrained exchange-correlation functional
for the unpolarized case in a way that is accurate for low densities.
Since the QMC calculations have lower variance for the fully polarized system,
we define $\epsilon^{\rm con}_{xc}(r_s,0) $ to be equal to the
difference between the fully polarized and unpolarized
energies of the antiferromagnetic spin chain with coupling constant $J(r_s)$,
determined using the WKB approximation.
Thus we rewrite the exchange-correlation functional for $\zeta=0$ as
\begin{equation}
  \fl \epsilon^{\rm con}_{xc}(r_s,0)  =  \frac{\epsilon_{xc}(r_s,0)}{1 + e^{\frac{r_s^2-O^2}{r_s R}}} +
    \left(1 - \frac{1}{1 + e^{\frac{r_s^2-O^2}{r_s R}}}\right) 
                                  \left(\epsilon_{xc}(r_s,1) - J(r_s)\ln 2 +
                                  \frac{\pi^2}{16 r_s^2}\right), \label{exc_z0con}
\end{equation}
where $O$ and $R$ are additional fitting parameters, and $J(r_s)$ is the same as in (\ref{jwkb})
with $J^\star=184.53$ and $\nu=2.84968$ determined via the WKB approach
for our potential in (\ref{potential}). In this way both the high and the low density limits are fulfilled.

Finally, the fully spin dependent density functional reads:
\begin{eqnarray}
\label{fullxcfit}
\fl \epsilon_{xc}(r_s, \zeta) = \epsilon^{\rm con}_{xc}(r_s,0)  + h_z(r_s, \zeta) + c_z(r_s, \zeta) +
\frac{1}{1 + e^{t(r_s)(1-|\zeta|)^\delta}} \\
\fl  \left[ 2 \left((1-w(r_s))\zeta^2 + w(r_s)\zeta^4 \right) 
        \left(\epsilon_{xc}(r_s,1) - \epsilon^{\rm con}_{xc}(r_s,0)\right) 
         - 2(h_z(r_s, \zeta) + c_z(r_s,\zeta))\right] \nonumber
\end{eqnarray}
where the additional functions are:
\begin{eqnarray}
t(r_s) =& \frac{t_1 e^{-t_2 r_s}}{r_s}, \\
w(r_s) =& e^{-w_1 r_s}, \\
c_z(r_s,\zeta) =& -C r_s^2 \zeta^2, \\
h_z(r_s,\zeta) =& \frac{r_s\ln\left(1 - (|\zeta|-h_\textrm{corr}(r_s,\zeta))^2\right)}{\pi^2 b^2}, \\
h_\textrm{corr}(r_s,\zeta) =& H_1 r_s^{H_2} \exp({-H_3 r_s}) \zeta^4 .
\label{fullxcfit-eqs}
\end{eqnarray}
$c_z(r_s,\zeta)$ is the small $\zeta$ expansion of the correlation energy around $\zeta=0$, while
$h_z(r_s,\zeta)$ is the variation of the exchange energy with respect to $\zeta=0$ at fixed $r_s$.
Both expressions are taken in the high density limit. $h_z(r_s,\zeta)$ includes another parametric
function ($h_\textrm{corr}(r_s,\zeta)$) which accounts for the non analytic behaviour of the exchange energy
around $\zeta=1$ at $r_s = 0$.
The form of (\ref{fullxcfit}) was chosen to constrain the
parametrization to attain energies determined by 
(\ref{exc_z0con}) and (\ref{exc_zeta}).  This allows the
parametrization to in principle satisfy the non analytic behaviour of the
correlation energy at high density and $\zeta=1$, while the low density behaviour
is fulfilled by the mapping onto the Heisenberg model. Even if the parametrization
looks complex at the first glance, there are only 21 independent parameters.

We have carried out extensive LRDMC simulations to find the best fitting parameters
for our parametrization. We note that there is another "external" parameter $b$, which
sets the effective thickness of the wire and therefore defines the interparticle potential.
It is of course possible to derive the parametrization
for different widths, but here
we chose to work with $b=1$, which is close to the usual thickness of wires realized
in semiconductor nanodevices.\cite{auslaender-science0}
The calculations for $b=1$ yield a series of total energies as a
function of density and spin polarization.

Great care is taken to remove all biases in the LRDMC calculations of the energy.
The lattice space error is
removed by calculating the energy for different lattice spacings and
extrapolating the results with a quadratic fit in the lattice space.
Finite size effects are removed by calculating the energy for
several numbers of electrons and extrapolating the result to the
thermodynamic limit by fitting the data to the form
\begin{equation}
E(N) = E + \frac{c_2 \sqrt{\ln{N}}}{N^2} + \frac{c_1}{N^2},
\label{finite_size_errors}
\end{equation}
where $E$ is the energy extrapolated to the thermodynamic limit, $N$ is the
number of electrons in the calculation, and the
constants $c_1$ and $c_2$ are fitting parameters determining the size of the
one-body and two-body finite size corrections.
Additionally the number of
electrons $N$ is chosen in each calculation so that the number of electrons
in each spin species is odd, thus avoiding degeneracy effects.
The form in (\ref{finite_size_errors}) is
obtained by following the finite size analysis described in Ref.~\cite{chiesa-finite-size}.

\begin{figure}[!ht]
\centering
\includegraphics[angle=-90,width=10cm]{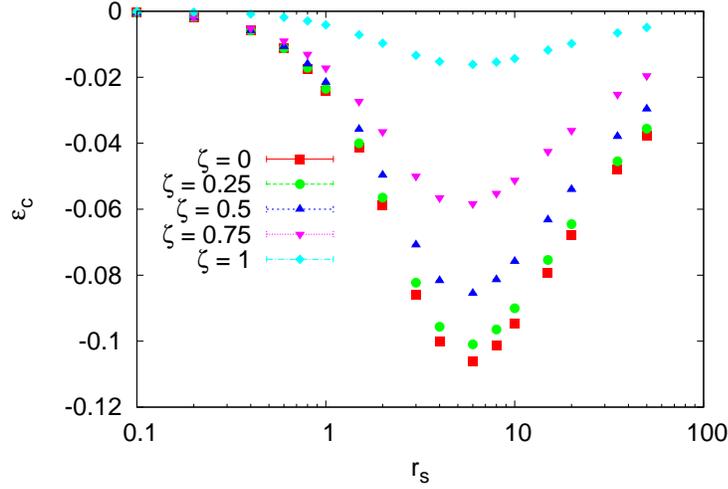}
\caption{
The correlation energy of the electron gas as a function of the density
$r_s$ is plotted for $b = 1$ at five values of the polarization, $\zeta$.
}
\label{b1rsenergies}
\end{figure}

\begin{figure}[!ht]
\centering
\includegraphics[angle=-90,width=10cm]{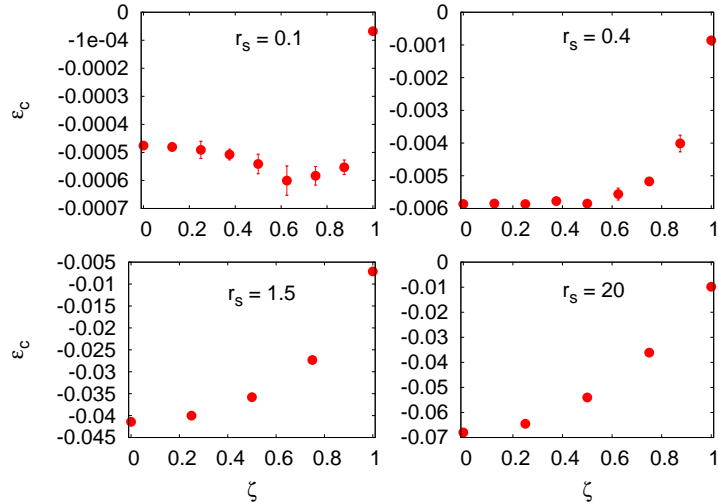}
\caption{
The correlation energy of the electron gas as a function of the polarization
$\zeta$ is plotted for $b = 1$ at four values of the density, $r_s$.
The upper right panel shows evidence of the RPA behaviour, quadratically as
a function of $\zeta$ near $\zeta = 0$ and then assuming an abrupt change
around $\zeta = 1$.  The upper right and lower left panels plot the
intermediate case, while
the lower right panel shows the $\zeta$ dependence at low density.}
\label{b1zetaenergies}
\end{figure}

Our results are plotted in Fig.~\ref{b1rsenergies} and Fig.~\ref{b1zetaenergies} which
show the behaviour of the correlation energy as a function of the density and
the polarization respectively.
The correlation energy at high density ($r_s = 0.1$) as a function of the
polarization shows vestiges of the non-analyticity in the correlation energy
at $\zeta = 1$ for $r_s \rightarrow 0$ (see (\ref{RPA_full1})).

\begin{figure}[!ht]
\centering
\includegraphics[width=10cm]{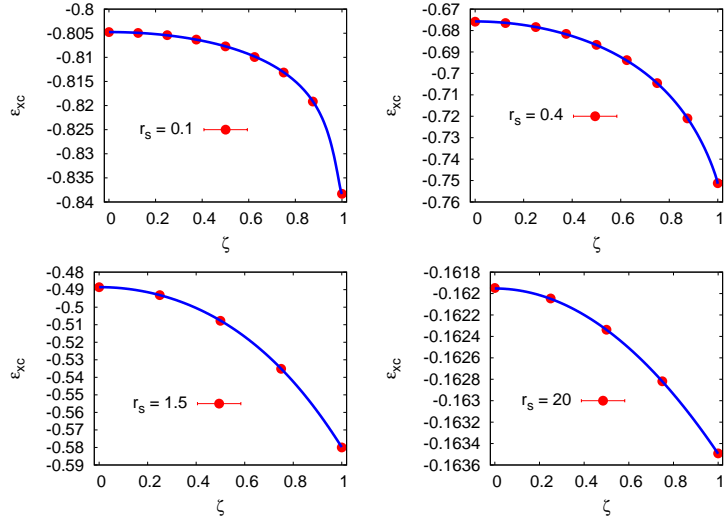}
\caption[Exchange-correlation Energy vs $\zeta$ at High Density for Exchange Correlation
Functional]{
Exchange-correlation energy $\epsilon_{xc}$ vs the polarization $\zeta$ at various densities.
The solid line comes from the parametrization while the points come from
QMC calculations. Their error bars are smaller than the point size.}
\label{xc_param_highdens}
\end{figure}

Tables \ref{tab1} and \ref{tab2} present the various parameters that are obtained by
a least-squares minimization fitting of the LRDMC values for the exchange-correlation energy 
computed at 17 different densities ranging from $r_s = 0.1$ to $r_s = 50$. From $r_s = 0.1$ to $r_s
= 1.5$ nine values of the polarization were used equally spaced from $\zeta
= 0$ to $\zeta = 1$.  For $r_s > 1.5$, five polarizations $\zeta = 0, \frac{1}{4}, \frac{1}{2},
\frac{3}{4}$, and $1$ were used.  These parameters produce a fit that has a
reduced $\chi^2$ of $5.3$ and an overall accuracy on the order of $10^{-5} Ryd^\star$.
The exchange correlation energy is
plotted at several values of the density in Fig.~\ref{xc_param_highdens}.
As one can see, it is in a good agreement with
the parametrization at all densities.

\begin{table}[!htp]
\centering
\begin{tabular}{|r|c||r|c|  }
\hline
$a_0$ &  -0.8862269 & $a_1$ & -0.8862269   \\
\hline
$b_0$ &  -2.1414101 & $b_1$ & -0.3326405   \\
\hline
$c_0$ &   0.4721355 & $c_1$ & -0.1771497   \\
\hline
$d_0$ &   2.81423    & $d_1$ & 0.653545   \\
\hline
$e_0$ &    0.529891  & $e_1$ & 0.374563  \\
\hline
$f_0$ &   0.458513  & $f_1$ &  0.171205 \\
\hline
$g_0$ &  -0.202642 & $g_1$ &  -0.101321 \\
\hline
$h_0$ &   0.470876 & $h_1$ &  0.281659 \\
\hline
$\alpha_0$ &  0.104435  & $\alpha_1$ & 0.097434  \\
\hline
$\beta_0$ &  4.11613 & $\beta_1$ &  2.86885  \\
\hline
$R$ &   1.25764   & & \\
\hline
$O$ &   3.11828   &  & \\
\hline
\end{tabular}
\caption{Parameter Values for the Fit of $\epsilon_{xc}^{con}(r_s, 0)$ and $\epsilon_{xc}(r_s, 1)$
\label{tab1}}
\end{table}

\begin{table}[!htp]
\centering
\begin{tabular}{|r|c|| r | c | }
\hline
$t_1$ &  2.31555  & $H_1$ &  5.90407  \\
\hline
$t_2$ &  1.83481 & $H_2$ &  2.44223 \\
\hline
$w_1$ &  0.83862 & $H_3$ &  2.93455  \\
\hline
$\delta$   & 0.70584    &  &    \\
\hline
\end{tabular}
\caption{Other parameters of the parametrization
\label{tab2}}
\end{table}

\section{Conclusions}
\label{conclusions}

In this paper we have presented results for properties of 
a quasi-one-dimensional electron gas,
with harmonic transverse confinement and long-range $1/r$ interactions,
which is a model for confined semiconductor structures.
By carrying out extensive lattice regularized diffusion Monte Carlo calculations,
we have determined the ground state energy as a function of spin and density,
and we have presented a parameterized fit to the Monte Carlo
data that can be used as a local density functional for exchange and correlation
in spin density functional calculations. 
The form is given in (\ref{fullxcfit}-\ref{fullxcfit-eqs}). It fulfills 
the high density limits of both exchange and correlation energies 
around $\zeta=0$ and at $\zeta=1$. At low density a mapping to an Heisenberg spin chain 
has been used to work out the $\zeta$ dependence, while the determination of $J$ comes from 
WKB calculations. The parameters are given in Tables \ref{tab1} and \ref{tab2} 
for a typical wire width $b=1$. The overall accuracy of the fit is on the order of $10^{-5} Ryd^\star$.

\ack
We thank D. M. Ceperley for useful
discussions, and Vinayak Garg for his careful reading of the manuscript.
 L.S., M.C., and R.M.M. acknowledge support in the form of the
NSF grant DMR-0404853.

\appendix

\section{RPA calculation of the spin dependent correlation energy}
\label{RPA_app}
In this appendix we compute the correlation energy
of a spin polarized 1DEG in the high density limit,
using the random phase approximation (RPA).
We start from the general expression of the RPA correlation energy\cite{pines-book}:
\begin{eqnarray}
\label{RPA_corr}
\fl \epsilon^{RPA}_{c}  = & \frac{L}{2 \pi} \int^{+\infty}_{-\infty}
dk~\epsilon(k),
\nonumber \\
\fl \epsilon(k) = & \frac{1}{4 \pi} \frac{|k|}{N} \int^{+\infty}_{-\infty} d\lambda ~
\ln( 1 - \tilde{v}(kb) \chi^0(k,ik\lambda) ) + \tilde{v}(kb) \chi^0(k,ik\lambda),
\end{eqnarray}
where $\tilde{v}(kb)$ is the Fourier transform of the potential,
and $\chi^0=\chi^0_\downarrow+\chi^0_\downarrow$ is the
real part of the density-density response function for the free 1D electron
gas:
\begin{equation}
\label{response}
\chi^0_\sigma(k,\omega) =  \frac{1}{4 \pi k} \ln\left(
\frac{\omega^2-(k^2-v^\sigma_F k)^2}{\omega^2 - (k^2+v^\sigma_F k)^2} \right),
\end{equation}
with $v^\sigma_F$ the Fermi velocity of the $\sigma$ ($=\uparrow,\downarrow$)
component.
After some algebra, and a change of variables ($k=k_Fq$,$\omega=ik_Fq
v_F u$), (\ref{RPA_corr}) can be rewritten at the leading $r_s$ order as
follows:
\begin{equation}
\label{RPA_approx}
\fl \epsilon^{RPA}_{c}  \simeq -\frac{1}{8(2 \pi)^3}
\int^{+\infty}_0 dq~q~\tilde{v}^2\left(\frac{qb}{\alpha r_s}\right) 
\int^{+\infty}_0 du (Q^\uparrow_q(u)+Q^\downarrow_q(u))^2,
\end{equation}
with $\alpha = 4 / \pi$ in 1D.
The derivation
follows the work
of Gell-Mann and Brueckner \cite{gellmann} in 3D,
and Rajagopal and Kimball \cite{rajagopal} in 2D. The
``propagator'' $Q^\sigma_q(u)$ depends now on the spin polarization,
and reads:
\begin{equation}
\fl Q^\sigma_q(u)  =  \int^{+\infty}_{-\infty} dk ~ \int^{+\infty}_{-\infty} dt ~
f_\sigma(k) (1-f_\sigma(k+q)) e^{-ituq} \exp(-|t| (\frac{1}{2} q^2 + kq)) ,
\end{equation}
where $f_\uparrow(x)=\theta(|x|-(1+\zeta)),
f_\downarrow(x)=\theta(|x|-(1-\zeta))$ are the zero temperature Fermi
distributions for the two spin components, $\theta$ being the step function:
\begin{equation}
\theta(x) = \left\{
\begin{array}{ll}
1 & \textrm{if $x<0$}\\
0 & \textrm{if $x \ge 0$}.
\end{array} \right.
\end{equation}
In order to factor out explicitly the $r_s$ order dependence in
(\ref{RPA_approx}),
we apply another change of variables, by rescaling $q$
($q \rightarrow \frac{\alpha r_s}{b} q$), and we integrate over
$u$. After these  steps, the RPA correlation energy reads:
\begin{equation}
\label{RPA_approx_2}
\fl \epsilon^{RPA}_{c}(r_s,\zeta)  \simeq  -\frac{1}{8 (2 \pi)^3}
 \left(\frac{\alpha r_s}{b}\right)^2 \int_0^{+\infty} dz~z~\tilde{v}^2(z)
\sum_{\sigma,\sigma^\prime} F_{\sigma,\sigma^\prime}\left(\frac{\alpha r_s}{b} z,
 \zeta \right),
\end{equation}
where we have defined the set of functions:
\begin{eqnarray}
\fl F_{\sigma,\sigma^\prime}(q,\zeta)=   \frac{2 \pi}{q} \int_{-\infty}^{+\infty} 
dk_1 f_\sigma(k_1)(1-f_\sigma(k_1+q)) \int_{-\infty}^{+\infty} dk_2
f_{\sigma^\prime}(k_2) (1-f_{\sigma^\prime}(k_2+q)) \times  \nonumber \\\frac{1}{q^2+q(k_1+k_2)},
\end{eqnarray}
where the $\zeta$  dependence is included in the zero temperature Fermi
distributions $f_\sigma(k)$. From the above equation it is apparent that
$F_{\downarrow,\uparrow}=F_{\uparrow,\downarrow}$.

For $\zeta=1$, $F_{\uparrow,\uparrow}(q,1) \ne 0$, while
$F_{\downarrow,\downarrow}(q,1)=F_{\uparrow,\downarrow}(q,1)=0~ \forall q$.
Since $F_{\uparrow,\uparrow}(0,1)=\pi/2$, for the fully polarized 1DEG
we obtain:
\begin{equation}
\epsilon^{RPA}_{c}(r_s,\zeta=1) = - \frac{A}{8 \pi^4 b^2} r_s^2,
\label{RPA_polarized}
\end{equation}
a result which is in agreement with
the mean spherical approximation.\cite{gold-gs-osci}

To evaluate $E^{RPA}_{c}(r_s,\zeta)$ at intermediate polarizations,
we need to compute the limits:
\begin{equation}
\lim_{x \rightarrow 0} F_{\sigma,\sigma^\prime}(x,\zeta) ~~~~\textrm{with
  $\zeta <  1$}.
\end{equation}
It turns out that $F_{\uparrow,\uparrow}(0,\zeta)=\pi/(1+\zeta)$,
$F_{\uparrow,\downarrow}(0,\zeta)=\pi$, and
$F_{\downarrow,\downarrow}(0,\zeta)=\pi/(1-\zeta)$. Thus, our  final result
for the spin dependent RPA correlation energy is the following:
\begin{equation}
\epsilon^{RPA}_{c}(r_s,\zeta) = \left \{
\begin{array}{ll}
- \frac{A}{2 \pi^4 b^2}(1+\frac{1}{1-\zeta^2}) r_s^2  & \textrm{if $ r_s\ll 2(1-\zeta)b/\alpha$} \\
- \frac{A}{8 \pi^4 b^2} r_s^2 & \textrm{if $\zeta=1$}
\end{array}
\right.
\label{RPA_full}
\end{equation}
Notice that when  $\zeta=0$ we recover the RPA correlation energy for the
unpolarized 1DEG derived in Ref.~\cite{casula-1d}.

\section*{References}
\bibliographystyle{unsrt}
\bibliography{1d_proceedings}

\end{document}